\documentclass[fleqn,10pt]{wlscirep}

\usepackage{amsmath}
\usepackage{float}
\usepackage{soul}

\usepackage{enumitem}
\usepackage[utf8]{inputenc}
\usepackage[T1]{fontenc}
\usepackage{nicefrac}
\usepackage{siunitx}
\DeclareSIUnit{\bits}{bits}
\usepackage[justification=justified,singlelinecheck=false]{caption}
\usepackage{braket}

\title{Time-bin encoded quantum key distribution over \SI{120}{\kilo\meter} with a telecom quantum dot source}

\author[1]{Jipeng Wang}
\author[1]{Joscha Hanel}
\author[1]{Zenghui Jiang}
\author[2]{Raphael Joos}
\author[2]{Michael Jetter}
\author[1]{Eddy Patrick Rugeramigabo}
\author[2]{Simone Luca Portalupi}
\author[2]{Peter Michler}
\author[3]{Xiao-Yu Cao}
\author[3,4]{Hua-Lei Yin}
\author[5]{Lei Shan}
\author[1,*]{Jingzhong Yang}
\author[1, 6]{Michael Zopf}
\author[1, 6]{Fei Ding}

\affil[1]{Institut f{\"u}r Festk{\"o}rperphysik, Leibniz Universit{\"a}t Hannover, Appelstra{\ss}e~2, 30167~Hannover, Germany}
\affil[2]{Institut f{\"u}r Halbleiteroptik und Funktionelle Grenzfl{\"a}chen, Center for Integrated Quantum Science and Technology (IQ\textsuperscript{ST}) and SCoPE, University of Stuttgart, Allmandring 3, 70569 Stuttgart, Germany}
\affil[3]{National Laboratory of Solid State Microstructures and School of Physics, Collaborative Innovation Center of Advanced Microstructures, Nanjing University, 210093 Nanjing, China}
\affil[4]{Department of Physics and Beijing Key Laboratory of Opto-electronic Functional Materials and Micro-nano Devices, Key Laboratory of Quantum State Construction and Manipulation (Ministry of Education), Renmin University of China, 100872 Beijing, China}
\affil[5]{Information Materials and Intelligent Sensing Laboratory of Anhui Province, Institutes of Physical Science and Information Technology, Anhui University, 230601 Hefei, China}
\affil[6]{Laboratorium f{\"u}r Nano- und Quantenengineering, Leibniz Universit{\"a}t Hannover, Schneiderberg 39, 30167 Hannover, Germany}

\affil[*]{jingzhong.yang@fkp.uni-hannover.de}

\begin{document}

\begin{abstract}
{Quantum key distribution (QKD) with deterministic single photon sources has been demonstrated over intercity fiber and free-space channels. The previous implementations relied mainly on polarization encoding schemes, which are susceptible to birefringence, polarization-mode dispersion and polarization-dependent loss in practical fiber networks. In contrast, time-bin encoding offers inherent robustness and has been widely adopted in mature QKD systems using weak coherent laser pulses. However, its feasibility in conjunction with a deterministic single-photon source has not yet been experimentally demonstrated. In this work, we construct a time-bin encoded QKD system employing a high-brightness quantum dot (QD) single-photon source operating at telecom wavelength. Our proof-of-concept experiment successfully demonstrates the possibility of secure key distribution over fiber link of \SI{120}{\kilo\meter}, while maintaining extraordinary long-term stability over 6 hours of continuous operation, that is highest secure key rate among the time-bin QKDs based on single-photon sources. This work provides the first experimental validation of integrating a QD single-photon source with time-bin encoding in a telecom-band QKD system. This development signifies a substantial advancement in the establishment of a robust and scalable QKD network based on solid-state single-photon technology.}    
\end{abstract}
\flushbottom
\maketitle

\section*{Introduction}

Quantum key distribution (QKD) offers a practical approach to realize physical-level confidentiality for the sharing of secret keys in a communication network~\cite{Shor2000,Li2015,Lo2005a,Scarani2008}. Since the first BB84 protocol~\cite{Bennett1984}, significant progress has been made to bridge the gap between the theoretically unconditional security and practical implementation~\cite{Xu2020,Gisin2002,Zhang2018}. Among the proposed methods, the decoy-state protocol plays a crucial role in practical QKD systems~\cite{Lo2005,Wang2005,Lim2014,Ma2005}. Using weak coherent pulses (WCPs) in the decoy-state protocol enables a secure and cost-effective implementation. As a result, it has been widely adopted in national and commercial QKD networks~\cite{Liao2017,Ribezzo2023,Chen2021,Cao2022,Paraiso2021,Oesterling2012}. Despite its success, the performance of decoy-state QKD with WCPs as approximations to ideal single-photons remains fundamentally constrained. The probability of true single-photon emission is upper-bounded by the Poisson statistics of WCPs~\cite{Bozzio2022,Wang2008,Pousa2024}, and additional modulation processes required to implement the decoy protocol may introduce complexity and side-channel vulnerabilities~\cite{Yoshino2018,Huang2019,Huang2018,Trefilov2024}. These limitations have motivated the pursuit of genuine single-photon sources (SPSs) for QKD.

Semiconductor quantum dots (QDs) embedded in nanophotonic structures offer on-demand, high-purity single-photon emission with high efficiency~\cite{Ding2024,Rota2024,Yang2024a}. Recent works have demonstrated the feasibility of using QDs as SPSs in QKD systems, both over fiber~\cite{Takemoto2015,Morrison2023,Intallura2009,Intallura2007,Gao2022,Schimpf2021,Zahidy2024,Zhang2025} and free-space~\cite{Basset2023,Rau2014,Samaner2022} channels. In particular, telecom-band QDs with Purcell enhancement~\cite{Nawrath2023} can provide high-brightness photons suitable for intercity fiber communication~\cite{Yang2024}, making them promising candidates for integration into practical QKD systems. Most existing QD-based QKD demonstrations rely on polarization encoding~\cite{Heindel2012,Zahidy2024,AlJuboori2023}, but such schemes are highly sensitive to polarization-mode dispersion (PMD) and birefringence in optical fibers~\cite{Li2019,Rodimin2024,LucioMartinez2009,Agnesi2019,Ding2017}. {This makes QKD systems vulnerable to changes in the practical quantum channel caused by environmental factors, such as turbulence, temperature and vibrations. This necessitates active compensation. In contrast, time-bin encoding, where qubits are encoded in the temporal position of single photons, offers intrinsic stability against such channel fluctuations even without any complex compensation protocols} \cite{Konteli2025,Qiu2025}. While time-bin encoding has been widely demonstrated using coherent-state or entangled-photon sources~\cite{Yin2020a,Boaron2018,Tanaka2012,Tang2014,Tang2022,Jin2019}, its implementation with deterministic QD-based SPSs remains largely unexplored.

Pioneering studies have utilized QDs for phase-encoding QKD~\cite{Takemoto2015,Intallura2009}, where asymmetric Mach-Zehnder interferometers (AMZIs) are used to create time-bin-like phase states. However, in those cases, the time-bin is not directly employed for key generation, and long-term stability tests are lacking. Entanglement-based QKD systems have also been explored~\cite{Anderson2020,Yu2025,Fitzke2022,Jayakumar2014,Chen2018,Lee2019,KhodadadKashi2025}, but they require complex state preparation techniques and are less practical for compact deployments. To date, there has been no demonstration of a QKD system employing genuine time-bin-encoding with deterministic single photons from QDs, especially at long distances.

In this work, we present a self-stabilized, time-bin encoded QKD system based on a deterministic telecom-wavelength QD source. {This source, involving an epitaxial InGaAs/GaAs QD embedded in a circular Bragg grating photonic structure has been previously reported with a high-brightness single-photon emissions\mbox{\cite{Nawrath2023}}}. To minimize system complexity and loss, we adopt a single phase modulator for preparing three quantum states: two time-bin basis states ($|Z_0\rangle$, $|Z_1\rangle$) and one phase-basis state ($|X_0\rangle$), assuming $\ket{X_0}$ shares the same error rate as $\ket{X_1}$ in the conventional BB84 protocol \cite{Bacco2019,Boaron2018}.The system is operated continuously for 6 hours, highlighting the intrinsic robustness of the time-bin scheme enabled by the system including the Sagnac interferometer (SNI), active feedback control, etc. Finally, we achieved a secure key bits (SKBs) per pulse of $2 \times 10^{-7}$ over a \SI{120}{\kilo\meter} fiber spool. This result confirms the feasibility of integrating QD single-photon sources into stable and field-deployable time-bin QKD systems, marking an important step toward scalable, quantum-secure communication networks.

\section*{Results}
\subsection*{Overview of the experimental scheme}
\begin{figure}[ht]
\centering\includegraphics[width=\textwidth]{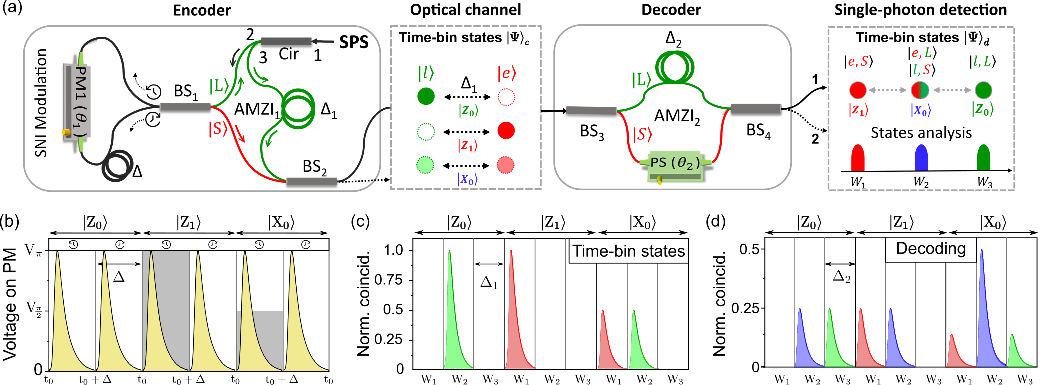}
\caption{\textbf{Encoding and decoding schemes of time-bin QKD. }(a) Encoder for preparing three time-bin states. Single photons emitted by the QD pass through the port 1$\to$2 of the optical circulator (Cir) and then through the beam splitter 1 ($BS_1$). A phase of $\theta_1$ in the set $\left \{ 0, \nicefrac{\pi}{2}, \pi \right \}$ is randomly encoded to the single photons via an electro-optic phase modulator (PM) within the SNI. Depending on the phase of single-photon interference, photons leave $BS_1$ via the short downward path or the long upward path (2$\to$3) to reach $BS_2$. Three single-photon path state $\ket{L}$ (green), $\nicefrac{1}{\sqrt{2}}\left ( \ket{S}+\ket{L} \right )$ and $\ket{S}$ (red) that are generated from the single-photon interference due to the phase $\theta_1$, are translated into three time-bin states $\left ( \ket{Z_0}, \ket{Z_1}, \ket{X_0} \right )$ after $BS_2$. After the transmission of the single photons through the optical channel, a phase shifter (PS) involved $AMZI_2$ at the decoder interpret the time-bin states to be measurable using the single-photon detection. The solid lines of $BS_2$ and $BS_4$ outputs denote the active port being used in the scheme{, while the inactive ports are specified as dashed arrow lines.} {Raw keys are sifted from the time windows $W_1$ (red), $W_2$ (blue) and $W_3$ (green) after base comparison between the encoder and decoder, respectively, corresponding to the encoded qubits $Z_1$, $Z_0$ and $X_0$.} (b) Sketch of the active phase control of single photons (yellow) via the PM in SNI configuration. Each single-photon period is divided into two time slots, considering the different arrival time of single photons at PM along superposition of clockwise $\left ( \circlearrowright \right)$ and counter-clockwise $\left ( \circlearrowleft \right)$ paths with a time delay of $\Delta$. One constant voltage in the set $\left \{ 0,V_{\pi}, V_{\nicefrac{\pi}{2}} \right \}$ (gray background) is applied to the PM to tune the phase of photons at the first time slot of each single-photon period, resulting in the generation of different path states from the SNI. (c) Sketch of single-photon correlation histograms with three time-bin states $\left ( \ket{Z_0}, \ket{Z_1}, \ket{X_0} \right )$ after the encoder. The time delay of $\Delta_1$ from $AMZI_1$ is revealed as the time delay between early $\ket{e}$ and late $\ket{l}$ photons . (d) {Sketch of the time-bin states correlation histograms from output 1 of the \mbox{$AMZI_2$} decoder. The histogram is normalised with the photon counts in (c).}}
\label{fig1_scheme}
\end{figure}

The three time-bin states of the polarized single photons are generated using an AMZI configuration involving a SNI. In this configuration, the input beam splitter of the standard AMZI is replaced with a fiber-based optical circulator (Cir), as shown in the left panel of Fig. \ref{fig1_scheme}a, so that the single photons are first guided into the SNI passing through $BS_1$. In the SNI structure, a $LiNbO_3$ phase modulator (PM, Rofea Optoelectronics, ROF-PM-UV) is intentionally placed in an unbalanced position. Single photons arriving at the PM along the clockwise $\left (\circlearrowright\right)$ path at the time $t_0+\Delta$ experience an additional time delay of $\Delta$ (half of the single-photon repetition period), compared to those arriving along the counterclockwise $\left (\circlearrowleft\right)$ path at the time $t_0$. 

In this experiment, we intentionally setup $\Delta$ to \SI{6.5}{\nano\second} as shown in Fig.\ref{fig1_scheme}b, considering the excitation repetition rate for the single photons is $f_{rep}=$\SI{75.947}{\MHz}. A correlation between the phase and the time of the photons arriving at the PM can be created by applying a sequence of two voltages to the PM within each single-photon period $\left ( \nicefrac{1}{f_{rep}}\approx\text{\SI{13.17}{\nano\second}} \right )$. 
{Within a single-photon period, a random voltage in the set $\left \{0,V_{\pi}, V_{\nicefrac{\pi}{2}} \right\}$ is first applied to the PM at $t_0$ for the first time slot, until no voltage is applied from $t_0+\Delta$ onwards for the second time slot. A random phase difference, $\theta_1$, can therefore be actively determined for each single photon between the $\circlearrowright$ and $\circlearrowleft$ paths. Subsequent single-photon interference at $BS_1$ leads to a superposition of the path state \mbox{\cite{Weihs2001}}, } 
\begin{equation}
    {
    \ket{\Phi}_{SNI} =  \left (-\sin\frac{\theta_1}{2} \cdot \ket{S}_{AMZI_1} + \cos\frac{\theta_1}{2} \cdot \ket{L}_{AMZI_1} \right )
    }
\end{equation}
The states $\ket{S}$ and $\ket{L}$ represent the quantum states of the short and long paths that the photons chose between the $BS_1$ and $BS_2$. The single photons with the state $\ket{L}$ enter the green path of the $AMZI_1$ and go through the Cir (through port 2$\rightarrow$3) in Fig. \ref{fig1_scheme}(a), giving a time delay of $\Delta_1$ in comparison with the state $\ket{S}$. This results in the equal time separation between the early $\ket{e}$ and late $\ket{l}$ photons after the $AMZI_1$. Assuming the transmitted and reflected single photons from the output port of $BS_2$ corresponds to the $\ket{L}$ and $\ket{S}$ single-photon states, the time-bin state of a photon from $AMZI_1$ is,
\begin{equation}
{
\begin{aligned}
     \ket{\Psi}_{c} &= \frac{1}{\sqrt{2}} e^{i\frac{\theta_1}{2}} \left (\sin\frac{\theta_1}{2} \cdot\ket{e} + i\cos\frac{\theta_1}{2} \cdot \ket{l} \right )\\
\end{aligned}}
\end{equation}
with $\nicefrac{1}{\sqrt{2}}$ indicating the amplitude of state from one port, and $i$ the phase shift of $\nicefrac{\pi}{2}$ for the  $\ket{S}$-state photons upon reflection relative to transmission at $BS_2$. In this work, a single output port is used for key transmission. Figure \ref{fig1_scheme}c shows the sketch of correlation histograms between the single-photon triggering signals and the three time-bin states  $\ket{Z_0}=\ket{l}$, $\ket{Z_1}=\ket{e}$ and $\ket{X_0}=\nicefrac{1}{\sqrt{2}}\left ( \ket{e} + i\ket{l} \right )$, corresponding to the voltage levels of $\left \{ 0, V_\pi, V_{\nicefrac{\pi}{2}}  \right \}$ shown in Fig. \ref{fig1_scheme}b, respectively. Within a single-photon period, three time windows $\left \{ W_1,W_2,W_3 \right\}$ are defined each with a range of $\Delta_1=$\SI{4.3}{\nano\second}. Coincidences that occur solely in $W_1$ and $W_2$ indicate the presence of the $\ket{e}$ and $\ket{l}$ photons, respectively. The probabilities of \SI{50}{\percent} for each $W_1$ and $W_2$ suggests the successfully encoded of $\ket{X_0}$ state. 

To decode the time-bin states, an AMZI with an internal time delay of $\Delta_2=\Delta_1$ and a phase shifter (PS, Luna Innovation, FPS-001) is employed. For simplicity, we ignore the global phase induced by the quantum channel in between the encoder and decoder, and exemplifying the phase $\theta_2 = \nicefrac{\pi}{2}$ from the PS. Then the single-photon state from output 1 of the $AMZI_2$ can be expressed as follows,

\begin{equation}
{
    \begin{aligned}
        \ket{\Psi}_{d} &=R_{AMZI_2} \cdot \ket{\Psi}_{c}\\
        &= \frac{1}{2\sqrt{2}}e^{i\frac{\theta_1}{2}} \left (  -i\sin\frac{\theta_1}{2}\ket{e}\ket{S}_{AMZI_2} + \sin\frac{\theta_1}{2} \ket{e}\ket{L}_{AMZI_2} + \cos\frac{\theta_1}{2}\ket{l}\ket{S}_{AMZI_2} + i\cos\frac{\theta_1}{2}\ket{l}\ket{L}_{AMZI_2}   \right) \\
    \end{aligned}}
    \label{equ:amzi2}
\end{equation}
where the $R_{AMZI_2}$ is operation gate of $AMZI_2$ for the single-photon state (see details in the methods). Here we assume that the phase shift of $\nicefrac{\pi}{2}$ is applied to single-photon states, when the $AMZI_2$ short path and active output port corresponds to the reflected photons from the $BS_3$ and $BS_4$, respectively. {Figure \mbox{\ref{fig1_scheme}}d represents the detection of a single photon at a set of given time windows \mbox{$\left \{ W_1,W_2,W_3 \right \}$} of output 1 corresponding to three cases,} 
\begin{itemize}[itemsep=0.1cm]
    \item {$W_1:$ The early photon goes through the short path $\ket{e,S}$;}
    \item {$W_2:$ The early photon goes through the long path $\ket{e,L}$ or the late photon goes through the short path $\ket{l,S}$; }
    \item {$W_3:$ The late photon goes through the long path $\ket{l,L}$;}
\end{itemize}

In the Equ. \ref{equ:amzi2}, the first term indicates the global phase induced by PM. {Meanwhile, the square of the coefficient for each term within the parentheses denotes the probability of each detected state.} {The sketch of correlation histograms in Fig. \mbox{\ref{fig1_scheme}}d illustrates the detection probability distribution of the above cases at output 1 of \mbox{$BS_4$} when the phase \mbox{$\theta_1$}, encoded by the PM, is set to \mbox{$\left \{ 0, V_{\pi}, V_{\nicefrac{\pi}{2}} \right \}$}. The $\ket{Z_0}$ state with late photonic qubits leads to photon detection at either $W_2$ or $W_3$, but only the photon in $W_3$ denote the $Z_0$ decoding basis. Likewise, early photonic qubits in $\ket{Z_1}$ states can be measured at $W_1$ or $W_2$ while the $W_1$ is the $\ket{Z_1}$ basis. For the encoded $\ket{X_0}$ qubits, when $\theta_1=V_{\nicefrac{\pi}{2}}$, the decoding basis is $W_2$ measured from output 1. This basis can be switched with that for $\ket{X_1}=\nicefrac{1}{\sqrt{2}}\left( \ket{e}-i\ket{l}\right )$, by controlling the phase difference of paths $\theta_2$ of $AMZI_2$. 

As with conventional QKDs using BB84 protocol, raw keys are sifted from the measured events according to shared basis information between users. In the time-bin-based QKD scheme, the decoder will gains the sifted keys by checking its measured results (the position of the detected event in time) based on the shared basis information from the encoder. For example, a '0' key will be sifted when the decoder learns the Z basis commonly used by the encoder and the photon is measured in $W_3$ (Fig. \mbox{\ref{fig1_scheme}}d). In analogy to the $\ket{Z_0}$ qubit, the raw keys '1' and '0' will be sifted when the common bases $\ket{Z_1}$ and $\ket{X_0}$ are revealed and the photons are detected within $W_1$ and $W_2$, respectively. In the following text, the bases of the \mbox{$\left\{ Z_0, Z_1, X_0 \right \}$} are coloured green, red and blue, to indicate the correlation with the time windows set that result in the sifted keys. For fig. \mbox{\ref{fig1_scheme}}d, note that for \mbox{$\ket{Z_0}$} and \mbox{$\ket{Z_1}$}, the same histograms will be measured at output 2, from which the other half of keys at the Z basis can be extracted from the time windows \mbox{$W_3$ and \mbox{$W_1$}}, respectively. However, due to the constructive interference $\left( \theta_2 = \nicefrac{\pi}{2} \right)$, the \mbox{$\ket{X_0}$} state can be analyzed directly using the detected event located within $W_2$ of output 1 (destructive interference pattern with the $W_2$ from output 2).}


\subsection*{Experimental setup}
\begin{figure}[ht]
\centering\includegraphics[width=\textwidth]{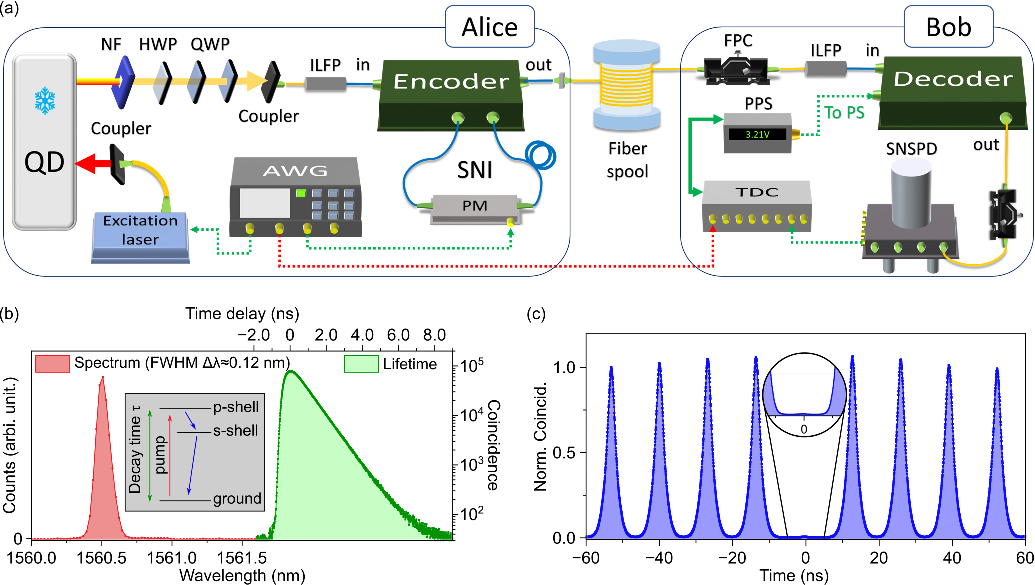}
\caption{\textbf{Experimental setups and source characteristics.} (a) Fiber-coupled excitation pulse laser, triggered by the arbitrary wave-function generator (AWG), transmits through a BS (R:T=98.5:1.5) and is used to excite the telecom QD loaded in the cryostat. {The emitted single photons collected by the objective are obtained by filtering out the laser using a notch filter.} Then, single photons are coupled to the in-line fiber polarizer (ILFP), through a series of free-space half- and quarter- waveplates (HWP and QWP) for the polarization control. The PM inside the encoder is synchronized with the excitation laser via the AWG, and randomly generate three single-photon time-bin states. The time-bin qubits are decrypted at the receiver setup from Bob, which consists of a decoder, superconducting nanowire single-photon detector (SNSPD), and a time-to-digital converter (TDC). {The excitation laser and PM at Alice are synchronized with the TDC at Bob via the AWG, which distributes electronic triggering signals through electrical cables.} FPC, fiber polarization controller; PPS, programmable power source. (b) The photoluminescence spectrum and time-resolved lifetime histogram (logarithm scale) of the single-photon emissions from the QD. The inset shows the decay process of the p-shell excitation within the QD band structure. (c) Normalized second-order autocorrelation histogram for the single photons from the trion state emissions with the inset showing the zoomed-in view of the central peak.}
\label{fig2_setup}
\end{figure}

{Figure \mbox{\ref{fig2_setup}}a shows a sketch of the experimental setup for time-bin QKD using telecom single photons from an InGaAs/GaAs QD involved in a circular Bragg grating photonic device reported in the previous work \mbox{\cite{Nawrath2023}}. {In this QKD system, Alice, acting as the sender, encrypts the time-bin states using single photons that are transmitted through the fibre spool to the receiver, Bob. Bob then performs decryption for the single-photon time-bin states. At Alice side,} the sample is loaded in a cryostat (Attodry 1100) at a temperature of \mbox{\SI{3.57}{\kelvin}}. A pulsed laser with a repetition rate of \mbox{$f_{rep}=$\SI{75.947}{\MHz}} synchronized with an arbitrary wave-function generator (AWG, Active Technologies, AWG5064) is used to excite the p-shell of the positive trion state of the QD. The single photon emissions with a central wavelength of \mbox{\SI{1560.6}{\nano\meter}} (Fig. \mbox{\ref{fig2_setup}}b) is collected by an objective with the numerical aperture of 0.7.} The total decay time is extracted by fitting the time-resolved QD emission in a three-level system and is found to be $\tau=$\SI{1018}{\pico\second}, {while the count drops to \mbox{\SI{1}{\percent}} up to \mbox{$\sim$\SI{4}{\nano\second}} . Taking into account the three time windows necessary for discriminating the time-bin states, we therefore apply the repetition rate \mbox{$f_{rep}=$\SI{75.947}{\MHz}} (corresponding to a window size of \mbox{\SI{4.3}{\nano\second}} for each). In a time-bin QKD system with a semiconductor single-photon source, employing the photonic resonant structure can reduce the lifetime and compromise the inherent limit of the repetition rates.}

{The encrypted raw time-bin key rate after a sequence of optical components and encoder is measured from the output of Alice approximately \mbox{\SI{162}{\kilo\Hz}}, involving the detection efficiency of \mbox{$\eta_D=$\SI{74}{\percent}}. This corresponds to average photon number per pulse of \mbox{$\left\langle n\right\rangle\approx2.89\times10^{-3}$} coupled to the quantum channel}. To evaluate the influence of the single-photon purity on QKD, we performed an autocorrelation measurement using a Hanbury Brown and Twiss setup and extract a blinking-corrected $g^{2}(0)=$\SI{0.85}{\percent} from the histogram (Fig. \ref{fig2_setup}c) without any temporal filter applied to the coincidence count integration $\left ( \tau =\nicefrac{1}{f_{rep}}\right )$. {Therefore, it is estimated that the upper bound of the multi-photon probability with a single-photon source is \mbox{$p_m\le\nicefrac{\left\langle n\right\rangle^2 \cdot g^{2}(0)}{2}$}. At short quantum channel length, the secure key rate (SKR) of QKD drops linearly with the quantum channel loss (logrithem scale), while the multi-photon probability limits the maximum tolerable loss significantly in the high-loss (long-distance) regime. In practical QKD, properly attenuating the single-photon rate can extend the MTL by compromising the raw key rate \mbox{$\left\langle n \right\rangle$} and the multiphoton portion \mbox{$p_m$} (detailed calculation in the Methods section).}

Before the time-bin encoder, the single photons are first polarized by an in-line fiber polarizer (ILFP) to align their polarization with the axis of the fiber optics, i.e., PM. In the SNI configuration, the phase control with the PM is implemented by an AWG that delivers squared modulation signals in pair with a clock rate locked to the excitation laser. Three voltage gaps $\left \{ 0, \text{\SI{1.6}{\volt}}, \text{\SI{3.2}{\volt}} \right \}$ corresponding to $\left \{ 0, V_{\frac{\pi}{2}}, V_{\pi} \right \}$ within each pair are applied to PM over the first time slot to generate the three time-bin states. In the actual experiment, a 16-bit repeating sequence with these random voltages are applied to the PM for the states of $\left \{ X_0,Z_1,Z_0,X_0,Z_0,Z_1,X_0,Z_1,Z_0,X_0,Z_0,Z_1,X_0,Z_0,Z_1,Z_1 \right \}$. { This leads to a basis choice ratio of \mbox{$\nicefrac{5}{16}$} and \mbox{$\nicefrac{11}{16}$} for X and Z basis, respectively, and number of bits \mbox{$\left \{5,6,5\right \}$} for \mbox{$\left \{ Z_0,Z_1,X_0 \right \}$}. It has been demonstrated that the asymmetric basis ratio in the BB84 protocol can improve the SKR \mbox{\cite{Morrison2023}}}. The encrypted single photons are then sent to the receiver setup through the variable-length fiber spools. Similar to the transmitter at Alice, the receiver at Bob uses an ILFP to ensure the alignment of the photon's polarization with the axis of the fiber optics in the decoder. A fiber polarization controller is placed in front to compensate the polarization drift from the fiber channel. Additionally, a programmable power source (PPS, Siglent Technologies, SPD3303) controls the PS in the decoder to actively stabilize the phase between the $AMZI_2$'s arms by minimizing the quantum bit error rate (QBER) of the system. Eventually, the arrival times of the single photons are registered by a superconducting nanowire single-photon detector (SNSPD), followed by a time-to-digital converter synchronized with the AWG. {The measured system dark count rate is \mbox{$d=\sim$\SI{100}{cts\per\second}}, resulting in the dark count probability \mbox{$p_{dc}=d\cdot\tau$}. This reduces the SKR and MTL, and is calculated in the Methods section.}

\subsection*{Evaluation of the QKD performance}
\begin{figure}[htbp]
\centering\includegraphics[width=\textwidth]{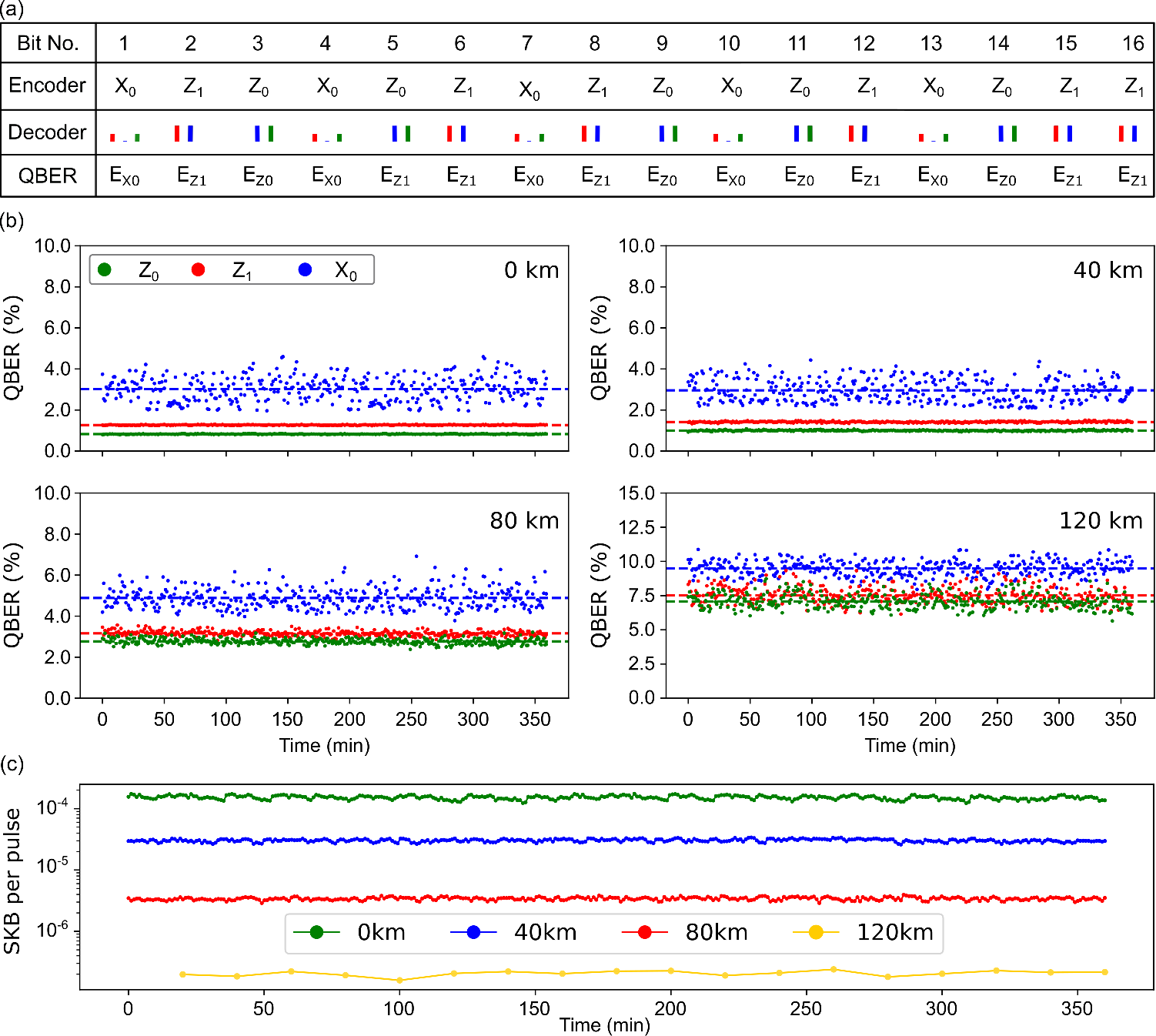}
\caption{\textbf{Time-dependent QKD performance at various transmission distances.} (a) Schematic of QBER extraction with the bases comparison of statistical data. The time windows $\left \{ W1, W2, W3 \right \}$ are indicated by the red, blue, and green columns, respectively. The height of each columns denotes the probability of detecting photons within a given time window when specific states are encoded. (b) Time-dependent QBER for different fiber spool lengths of \SI{0}{\kilo\meter}, \SI{40}{\kilo\meter}, \SI{80}{\kilo\meter}, and \SI{120}{\kilo\meter}. Each data point represents \SI{1}{\minute} of measurement time. The dashed lines indicate the average QBERs on the bases over the 6 hours. (c) Secure key bits (SKBs) per pulse as a function of measurement time for the different fiber spool lengths ranging from \SI{0}{\kilo\meter} to \SI{120}{\kilo\meter}. For the spool lengths of \SI{0}{\kilo\meter}, \SI{40}{\kilo\meter}, and \SI{80}{\kilo\meter}, an acquisition time of \SI{1}{\minute} $\left ( N_{sum} = 4.56 \times 10^9 \right )$ is used in the finite key analysis. For the \SI{120}{\kilo\meter} transmission distance, an acquisition time of \SI{20}{\minute} $\left ( N_{sum} = 9.12 \times 10^{10} \right )$ is used.  
}
\label{fig3_performance}
\end{figure}
As the figure-of-merit for QKDs, the SKR $\left( R_{secure} \right)$ from the three time-bin states, is emulated in the finite-key regime with the multiplicative Chernoff bound \cite{Morrison2023},
\begin{equation}
    R_{secure}=\left\lfloor \underline{N}^{Z}_{R,nmp}\left(1-h\left(\overline{\phi}_Z\right)\right)-\lambda_{EC}-2 \log _2 \frac{1}{2 \varepsilon_{PA}}-\log _2 \frac{2}{\varepsilon_{cor}}\right\rfloor/N_{sum}
\end{equation}
{In this process, the raw keys on the Z basis are post-processed to create secure keys after error correction and privacy amplification, while the raw keys from the X basis are shared publicly to analyse the system QBER. \mbox{$\underline{N}^{Z}_{nmp}$} denotes the lower bound of the received raw key rate on the Z basis, excluding the noisy bit rates caused by multi-photon emission that can be estimated based on second-order correlation $g^{(2)}(0)$. As it is impossible to obtain the QBER on the Z basis,  the upper bound of the phase error that is translated from the QBER on the X basis is employed. This is involved in, $h(\cdot)$ the binary Shannon entropy, accounting for {eavesdropper's} attack on the quantum raw keys}. $\lambda_{EC}$ is the lower bound of information leakage during the error correction. $\epsilon_{PA}$ and $\epsilon_{cor}$ are the security parameter over the error verification, and failure probability of privacy amplification, respectively. {In the finite-key regime, which closely resembles a practical scenario, the quantum keys are sent in blocks, with $N_{sum}$ specifying the size of the key block sent from the encoder.} Details about the system parameters and the calculation model are provided in the Methods section. 

{The QBERs are extracted from the histograms as presented in Fig. \mbox{\ref{fig3_performance}}a, in which the number of raw keys bits and error bits are counted. The decoder first measures the 16 histograms corresponding to the time-bin states repeatedly sent by the encoder.} The QBER for each basis is calculated using the ratio of the integrated photon counts at the two perpendicular bases. For instance, the QBER for encoded $E_{Z_1}$ is calculated as $\nicefrac{N_{Z_0}}{N_{Z_0}+N_{Z_1}}$, where $\left \{ N_{Z_0}, N_{Z_1} \right\}$ are the integrated photon counts within each \SI{4.3}{\nano\second} time windows $\left \{ W3,W1 \right\}$ of the histograms (Bit no. 2, 5, 8, 12, 15, 16). Same calculation algorithm applies to $E_{Z_0}$, which is $\nicefrac{N_{Z_1}}{N_{Z_0}+N_{Z_1}}$ with $\left \{ N_{Z_0}, N_{Z_1} \right\}$ counted within the time window $\left \{ W3,W1 \right\}$ from the corresponding histograms (Bit no. 3, 5, 9, 11, 14). However, determining $E_{X_0}$ is relatively challenging due to the absence of one more detector channel of $AMZI_2$ for the $N_{X_0}$ and $N_{X_1}$ at the same time. We adjust PS phase to be $-\nicefrac{\pi}{2}$ while sending the $\ket{X_0}$, such that theoretically there is no correlation peak within the $W_2$ window from the $\ket{X_0}$ state due to the destructive interference. Then, we regard the detected error qubits as $N_{X_1}$ similar to four-state BB84 protocol, and assume the QBER of $\ket{X_0}$ state to be $E_{X_0} =\nicefrac{N_{1}}{N_{Z_0}+N_{Z_1}}$, as the splitting ratio between X- and Z- basis at the decoder is $\nicefrac{1}{2}$ resulting in $N_{X_0}+N_{X_1}=N_{Z_0}+N_{Z_1}$ (Bit no. 1, 4, 7, 10, 13). For the measurement of $E_{X_0}$, the phase difference between the paths of $AMZI_2$ is dynamically stabilized by suppressing the photon counts $N_{X_1}$ within the time window $W_2$ to be the minimal.


To investigate the stability of the time-bin QKD system in terms of the QBER, {SKBs per pulse \mbox{$\left ( \nicefrac{R_{secure}}{f_{rep}} \right )$}} at different quantum channel lengths, we implement the time-dependent measurement using the length-variable fiber spools (average loss of $\alpha=$\SI{0.1956}{\dB\per\km}) connected between the transmitter and receiver setups. As shown in Fig. \ref{fig3_performance}b, {the mean and deviation of the QBERs at X basis are both relatively higher than Z basis. This is due to the limited visibility of interference at \mbox{$AMZI_2$} (i.e., imperfect power splitting ratio of the BSs), as the accurate detection of the \mbox{$\ket{X_0}$} requires high-quality single-photon interference at the \mbox{$BS_4$}. This is revealed by the higher misalignment probability of the optical setup on the X basis than the Z basis \mbox{$\left( p_{misX} >  p_{misZ} \right)$}, which contributes bit errors at a transmission distance of \mbox{\SI{0}{\kilo\meter}}.} We attribute the slightly higher QBERs of $\ket{Z_1}$ state compared to the $\ket{Z_0}$ state to the imperfect phase encoding of PM due to inaccurate targeting voltage (i.e., flatness uncertainty of the peak voltage). {As represented in Fig.\mbox{\ref{fig1_scheme}}b, an ideal \mbox{$\ket{Z_0}$} state can be generated without applying any voltage to the PM over a single-photon period. However, voltages \mbox{$V_{\pi}$} and \mbox{$V_\frac{\pi}{2}$} with time duration of \mbox{$\nicefrac{1}{2f_{rep}}$} is required to obtain \mbox{$\ket{Z_1}$} and \mbox{$\ket{X_0}$} state, respectively. The uncertainties in these voltages translates into the additional QBERs for the \mbox{$\ket{Z_1}$} and \mbox{$\ket{X_0}$}} states. On the other hand, the QBER increases with the length of the optical fiber, since the system dark counts become more dominant with a decreased signal-to-noise ratio. Nevertheless, average QBERs below \SI{11}{\percent} are maintained at a transmission distance of \SI{120}{\kilo\meter}, which is promising for a secure intercity-scale communication. {Fiber-induced light dispersion causes an elongation of the single-photon pulses for $\sim\SI{265}{\pico\second}$ at a transmission distance of \mbox{\SI{120}{\kilo\meter}} considering the linewidth of the emitted single photon is $\Delta\lambda\approx$\mbox{\SI{0.13}{\nano\meter}} \mbox{\cite{Nawrath2023}}. This gives a negligible influence on the time-bin qubits, with a time separation of \mbox{\SI{4.3}{\nano\second}}.} Figure \ref{fig3_performance}c illustrates stable SKB per pulse over 6 hours for different fiber spools, which is the ratio of $R_{secure}$ and $f_{rep}$. The average $E_{X_0}$ within the integration time is dynamically employed in the calculation of $R_{secure}$, while considering fixed values of the mean photon number per pulse $\left \langle n \right \rangle=2.89\times10^{-3}$ entering the quantum channel and $g^{(2)}(0)=$\SI{0.85}{\percent}. The integrated finite blocks at a distance of \SI{120}{\kilo\meter} are given a longer time of \SI{20}{\minute} to ensure sufficient key length for a positive key rate.
\begin{table}[ht]
\centering
\begin{tabular}{|c|c|c|c|c|c|c|c|}
\hline
\textbf{Distance (km)} & \textbf{$\nicefrac{R_{secure}}{f_{rep}}$} & \textbf{$\nicefrac{R_{raw}}{f_{rep}}$} & \textbf{$E_{Z}$ (\%)} & \textbf{$\sigma_{Z_0}$ (\%)} & \textbf{$\sigma_{Z_1}$ (\%)} & \textbf{$E_{X_0}$ (\%)} & \textbf{$\sigma_{X_0}$ (\%)}  \\
\hline
0 & $1.59 \times 10^{-4}$ & $2.23 \times 10^{-4}$ & 0.98\% & 0.01\% & 0.01\% & 3.14\% & 0.54\%  \\

40 & $3.04 \times 10^{-5}$ & $4.33 \times 10^{-5}$ & 1.19\% & 0.08\% & 0.03\% & 3.12\% & 0.56\%  \\

80 & $3.54 \times 10^{-6}$ & $6.87 \times 10^{-6}$ & 3.02\% & 0.13\% & 0.14\% & 4.90\% & 0.52\%  \\

120 & $1.99 \times 10^{-7}$ & $1.34 \times 10^{-6}$ & 6.85\% & 0.60\% & 0.56\% & 9.60\% & 0.58\%  \\
\hline
\end{tabular}
\caption{\textbf{SKB per pulse and QBERs over a range of fiber spool lengths.} The QBER on Z-basis, $E_Z $, is the average value of $E_{Z_0}$ and $E_{Z_1}$. Acquisition times of \SI{1}{\minute} and \SI{20}{\minute} with the key blocks in the finite key regime are used for \SI{0}{}$\sim$\SI{80}{\kilo\meter} and \SI{120}{\kilo\meter}, respectively.}
\label{tab:SKB_QBER}
\end{table}

\begin{figure}[H]
\centering\includegraphics[width=\textwidth]{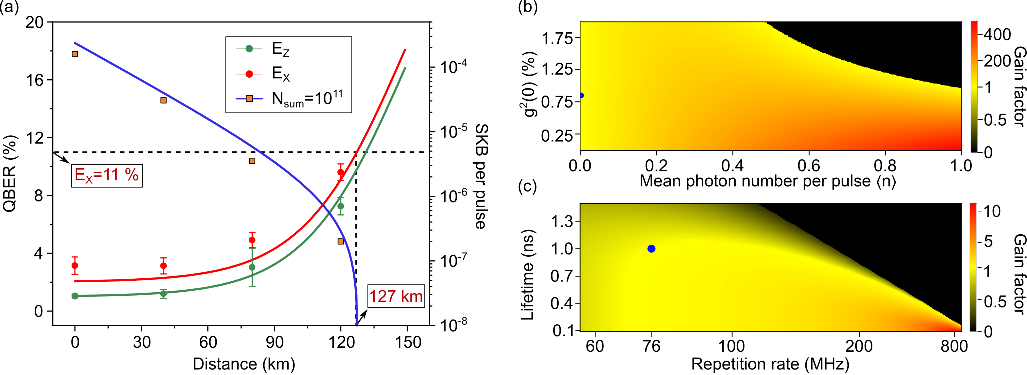}
\caption{\textbf{Time-bin QKD performance versus transmission distance and gain in secure key rate with improved source qualities.} (a) QBERs at the $\ket{X} \text{and} \ket{Z} $ bases and SKB per pulse as a function of the secret key transmission distance. A received block size of $N_{sum}=10^{11}$ is employed for the finite key analysis. (b) Gain of simulated SKR as a function of the mean photon number per pulse $\left \langle n \right \rangle $ and second order autocorrelation $g^{(2)}(0)$, compared with this experiment. (c) Gain of simulated SKR as a function of the repetition rate of the excitation laser and the QD lifetime, compared with this  experiment. The blue circles indicate the parameters in the current time-bin QKD system.
}
\label{fig4_key_vs_distance}
\end{figure}

Table \ref{tab:SKB_QBER} presents the statistics of the Gaussian distribution according to the results from Fig. \ref{fig3_performance}. The extraction ratio of the $R_{secure}$ from the raw key rate $R_{raw}$ becomes lower with the enhanced transmission distance because of the increased QBER. With the case of the repetition rate $f_{rep}=$\SI{75.947}{\MHz}, the reachable SKR at the distance of \SI{120}{\kilo\meter} is approx. \SI{15}{\bits\per\second}, which is still possible for the text message encryption. The standard deviation of the QBERs $\left ( \sigma_{\left \{ Z_0,Z_1,X_0 \right \} } \right )$ on both the Z- and X-bases remains below \SI{0.6}{\percent} to be constant thanks to the effective phase compensation program and stable laboratory environment. Fig. \ref{fig4_key_vs_distance}(a) presents the QBERs and the SKB per pulse as a function of the transmission distance. Apart from the experimental data points as illustrated in the table, we performed a simulation to determine the maximum tolerable distance for our time-bin QKD system, where the QBERs of $E_X$ and $E_Z$ on Z- and X-bases ($\phi^Z$ and $\phi^X$) are simulated as,
\begin{equation}
\centering
    E_X = \frac{M_R^X}{\underline{N}_{R,nmp}^X} \hspace{3cm} 
    E_Z = \frac{M_R^Z}{\underline{N}_{R,nmp}^Z}
\end{equation}

where $M_R^{X,Z}$ and $\underline{N}_{R,nmp}^{X,Z}$ denote the number of error bits and lower bound of non-multiphoton fraction of received photons at $\left \{ X,Z \right\}$ bases, respectively (see Methods for details). A maximum tolerable distance of \SI{127}{\kilo\meter} is underestimated in the case that the QBER at X-basis approaches \SI{11}{\percent}, since keys from Z-basis with a lower QBER value are typically employed for the information encryption in practical QKDs.

\section*{Discussion}
In our experiment, a secure key rate of $1.99 \times 10^{-7}$ bit per pulse was achieved over a \SI{120}{\kilo\meter} fiber spool using a total pulse block size of $N_{\text{sum}} = 10^{11}$, corresponding to an integration time of approximately 1300~s. This result demonstrates the feasibility of employing a deterministic, telecom-band QD single-photon source in a time-bin encoded QKD system under long-distance transmission and realistic conditions. Nevertheless, there remains considerable potential for further improvement in system performance, as discussed below:

\paragraph{1. Influence of source brightness and system loss.}
The mean photon number $\langle n \rangle$ is a critical parameter affecting the key rate. Increasing source brightness and reducing encoder loss can significantly enhance $\langle n \rangle$ compared to current experimental conditions. As shown in Fig.~\ref{fig4_key_vs_distance}(b), the SKR improves with larger $\langle n \rangle$. However, higher brightness increases sensitivity to multi-photon components, and a low $g^{(2)}(0)$ becomes increasingly important to preserve security. Poor single-photon purity (i.e., high $g^{(2)}(0)$) has a stronger negative impact at higher source brightness.

\paragraph{2. Repetition rate limitations imposed by QD lifetime and modulation structure.}
Compared with weak coherent laser pulses, QD sources typically exhibit longer radiative lifetimes. Therefore, increasing the system repetition rate leads to temporal overlap between $\circlearrowright$ and $\circlearrowleft$ wave packets within the Sagnac interferometer (SNI). This overlap region exhibits the same phase and thus lacks modulation contrast, making it unusable for key generation. Additionally, at higher repetition rates, overlap between detection windows (W1, W2, and W3) may occur, resulting in photons from adjacent bits falling into incorrect time bins and increasing the quantum bit error rate (QBER). To further explain this point, we performed simulations based on a fitted model of the QD lifetime to explore the trade-off between repetition rate and temporal overlap, as shown in Fig.~\ref{fig4_key_vs_distance}(c). While higher repetition rates can theoretically enhance the key rate, they are only effective when the pulse lifetime is sufficiently short to prevent peak overlap. An optimal operating point must balance increased repetition with minimal temporal crosstalk.

\paragraph{3. Optical loss and visibilities in the encoding and decoding modules.}
The secure key rate is also constrained by the intrinsic loss in both Alice’s and Bob’s modules. Several components in the system can be optimized further, such as using lower-loss fiber devices and replacing standard fiber connectors with high-precision fusion splicing, thereby minimizing insertion loss and back-reflection. {To further reduce the QBERs of the \mbox{$\ket{X_0}$} state, AMZIs with high- and stable visibilities needs to be carefully optimised by employing ultra-balanced BSs and automate feedback phase shift control.} 

\paragraph{4. Detector performance and dark count suppression.}
The performance of the SNSPD plays a crucial role in system reliability. Although the detectors used in this work exhibit good efficiency and low dark count probability, further improvements are possible. Enhancing detector efficiency and reducing background counts through improved device fabrication and environmental isolation could boost the overall key rate.

\paragraph{{5. Synchronization between encoder and decoder at remote sites.}} 
{Synchronization between Alice and Bob is essential for a working QKD system to enable basis comparison using timing information. In our laboratory implementation, this is achieved by distribution of electronic clock signals. For a real-world point-to-point QKD system, various methods have been proposed to realize synchronization without relying on additional hardware or external references such as GPS} \cite{Calderaro2020,Cochran2021,Krause2025,Spiess2023,Zahidy2023}.

In summary, we have demonstrated the feasibility and long-term self-stability of a time-bin encoded QKD system based on a deterministic single-photon source at telecom wavelengths. {Benefiting from a stable emission of high-brightness and pure telecom single photons \mbox{\cite{Nawrath2023}}, the system operates continuously for over 6 hours at \mbox{\SI{120}{\kilo\meter}} and achieves a highest finite-size key rate among the time-bin QKDs with single-photon sources.} Our work identifies key advantages and also limitations of QD single photon sources for the generation of time-bin qubits. The results provide practical paths for optimization of all system components, therefore contributing to the realization of a robust and scalable quantum communication infrastructure based on solid-state single-photon emitters. 


\section*{Materials and methods}
\subsection*{Transformation of quantum states with the AMZI}
For an asymmetric Mach-Zehnder interferometer (AMZI), consisting of two beam splitters and a phase shifter (fast axis along with H polarisation of single photons) for one arm, the transformation matrix for the single-photon states before the $BS_4$ is,
\begin{equation}
    \begin{aligned}
        R_{AMZI}^\prime & = R_{PS} \otimes R_{BS_3} \\
        & = \begin{pmatrix}  e^{i\theta_2} & 0\\ 0 & 1 \end{pmatrix}_{PS} \otimes \frac{1}{\sqrt{2}} \begin{pmatrix} i & 1\\ 1 & i\end{pmatrix}_{BS_3}
    \end{aligned}
\end{equation}
where the $R_{BS}$, $R_{PS}$ are the transformation matrix for the BS and PS, respectively. Taking into account the single-photon state from the quantum channel is,
\begin{equation}
    \begin{aligned}
        \ket{\Psi}_{c} &= \frac{1}{\sqrt{2}} e^{i\frac{\theta_1}{2}} \left (\sin\frac{\theta_1}{2} \cdot\ket{e} + i\cos\frac{\theta_1}{2} \cdot \ket{l} \right )\\
        &=\frac{1}{\sqrt{2}}  e^{i\frac{\theta_1}{2}} \cdot
        \begin{pmatrix}  \sin\frac{\theta_1}{2} \\ i\cos\frac{\theta_1}{2} \end{pmatrix}_{T} \otimes \begin{pmatrix} 1 \\0  \end{pmatrix}_{P}
    \end{aligned}
\end{equation}
with the first two terms the time-bin states $T$ for the $\ket{e}$ and $\ket{l}$ photons from the quantum channel. The third term indicates the path states $P$ corresponding to $\ket{S}\text{ and} \ket{L} $ before $AMIZ_2$. The single-photon states before $BS_4$ is then,
\begin{equation}
    \begin{aligned}
        \ket{\Psi}^\prime_{AMZI_2} &= R^\prime_{AMIZ} \cdot \ket{\Psi}_{1} \\
        &= \frac{1}{\sqrt{2}}  e^{i\frac{\theta_1}{2}} \cdot
        \begin{pmatrix}  \sin\frac{\theta_1}{2} \\ i\cos\frac{\theta_1}{2} \end{pmatrix}_{T} \otimes \frac{1}{\sqrt{2}} \begin{pmatrix} ie^{i\theta_2} \\ 1 \end{pmatrix}_{P}\\
        & = \frac{1}{2}  e^{i\frac{\theta_1}{2}} \cdot \left ( i\sin\frac{\theta_1}{2}e^{i\theta_2}\cdot\ket{e}\ket{S}_{AMZI_2}+\sin{\frac{\theta_1}{2}}\cdot\ket{e}\ket{L}_{AMZI_2}-\cos{\frac{\theta_1}{2}}e^{i\theta_2}\cdot\ket{l}\ket{S}_{AMZI_2}+i\cos{\frac{\theta_1}{2}}\cdot\ket{l}\ket{L}_{AMZI_2} \right)
    \end{aligned}
\end{equation}
In our experiment, only one output port of $BS_4$ is used for the measurement. The final state from one port of the $AMZI_2$ can be written as follows, with a amplitude factor of $\nicefrac{1}{\sqrt{2}}$ is applied considering the splitting probability of 50:50 with the BS. In addition, assuming that the output of $BS_4$ corresponds to photons from the short path, it leads to a phase shift of $\nicefrac{\pi}{2}$ with the $\ket{S}$-state photons sigified by i. 
\begin{equation}
    \begin{aligned}
        \ket{\Psi}^\prime_{AMZI_2}&\rightarrow\ket{\Psi}_{d}\\
        &= \frac{1}{2\sqrt{2}}  e^{i\frac{\theta_1}{2}} \cdot \left ( -\sin\frac{\theta_1}{2}e^{i\theta_2}\cdot\ket{e}\ket{S}_{AMZI_2}+\sin{\frac{\theta_1}{2}}\cdot\ket{e}\ket{L}_{AMZI_2}-i\cos{\frac{\theta_1}{2}}e^{i\theta_2}\cdot\ket{l}\ket{S}_{AMZI_2}+i\cos{\frac{\theta_1}{2}}\cdot\ket{l}\ket{L}_{AMZI_2} \right)
    \end{aligned}
\end{equation}

\subsection*{Estimation of QBERs}
In our time-bin QKD system, we estimate the QBERs and SKRs based on the calculation of click $p_{click}^{X,Z}$ and error $p_{e}^{X,Z}$ probability with the detected photons by SNSPD at $\left \{ X, Z\right \}$ bases, taking into account of the system parameters such as mean photon number per pulse $\left \langle n \right \rangle$, $g^{(2)}(0)$, total system loss (incl. fiber spools) $\eta_{total}$, etc.

\begin{equation}
 \begin{aligned}
    p_{c}^{X,Z} &=\sum_{n=0}^{\infty} p_n [ 1 - (1 - p_{dc}) (1 - \eta_{total})^n ] \\
    p_{e}^{X,Z} &= p_0 p_{dc} + \sum_{n=1}^{\infty} p_n \left[ 1 - (1 - p_{dc}) (1 - \eta_{total})^n \right] p_{mis}
\end{aligned}   
\end{equation}
in which $p_{dc}$ is dark count probability equal to the multiplication of system dark counts $d$ and individual time window $\tau_W=$\SI{4.3}{\nano\second}. Here, the parameter $p_{mis}^{X, Z}$ is the error probability of the signal components due to imperfect state preparation, channel decoherence, and imperfect power splitting at decoder. This is given by the average QBER for an optical fiber length of \SI{0}{\kilo\meter} in the experiment. The probability of n-photon emission $p_n$ is calculated as \cite{Morrison2023},
\begin{equation}
p_2 = \frac{\Bar{n}^2 \cdot g^{(2)}(0)}{2}
\qquad
p_1 =\Bar{n} - 2p_2
\qquad
p_0 = 1 - p_1 - p_2
\end{equation}
with $\Bar{n} = \left\langle n \right\rangle \cdot \eta_B \cdot \eta_D$ the average photon number per pulse after the detector. In the simulation of Fig. \ref{fig4_key_vs_distance} about the QBER as a function of transmission distance, we employ the phase error rate to estimate the QBER in the finite key length regime,
\begin{equation}
\centering
    E_X = \phi^Z = \frac{M_R^X}{\underline{N}_{R,nmp}^X} \hspace{3cm} 
    E_Z = \phi^X = \frac{M_R^Z}{\underline{N}_{R,nmp}^Z}
\end{equation}
in which $M_R^{X,Z}$ and $\underline{N}_{R,nmp}^X$ are calculated as,
\begin{equation}
\begin{aligned}
M_{R}^{X,Z} &= N_{sum}\cdot p_{X,Z}^{A} \cdot p_{X,Z}^{B} \cdot p_{e}^{X,Z} \\
\underline{N}_{R,nmp}^X &= N_{R}^{X,Z} - \Bar{N}_{sum,mp}^{X,Z}\\
& = N_{sum}\cdot p_{X,Z}^{A} \cdot p_{X,Z}^{B} \cdot p_{c}^{X,Z} - N_{sum}\cdot p_{X,Z}^{A} \cdot p_{X,Z}^{B} \cdot p_m
\end{aligned}
\end{equation}
$p_{X,Z}^{A,B}$ is the splitting ratio of the keys for the $\left \{ X,Z \right\}$ bases at the encoder and decoder sites. $p_m$ is the the multi-photon emission probability of the source, which is equal to $p_2$ in our case by only taking into account the multi-photon events up to two. $\Bar{N}_{sum,mp}^{X,Z}$ denote the upper bound of the emitted photons from the encoder that is derived with the upper Chernoff bound and $N_{sum,mp}^{X,Z}$, 
\begin{equation}
    \Bar{x} = (1+\delta_U)x^*
\end{equation}
with $\delta^U = \frac{\beta + \sqrt{8\beta x^* + \beta^2}}{2x^*}$ and $\beta = -\log_e (\epsilon_{PE})$.
\subsection*{Calculation of SKR}
The calculation of SKR in finite key regime based on the Chernoff bound has been discussed in the previous publications for the polarization-encoded QKDs \cite{Morrison2023,Yang2024},
\begin{equation}
    R_{secure}=\left\lfloor \underline{N}^{Z}_{R,nmp}\left(1-h\left(\overline{\phi}_Z\right)\right)-\lambda_{EC}-2 \log _2 \frac{1}{2 \varepsilon_{PA}}-\log _2 \frac{2}{\varepsilon_{cor}}\right\rfloor/N_{sum}
\end{equation}
with $\Bar{\phi}_Z$ calculated as,
\begin{equation}
    \begin{aligned}
        &\overline{\phi_{Z}} = \phi_{Z} + \gamma^U \left( N_{R,nmp}^{X}, N_{R,nmp}^{Z}, \phi_{Z}, \frac{\varepsilon_{sec}}{6} \right), \hspace{2cm} \phi_{Z}=E_{X}=\frac{M_{R}^{X}}{\underline{N_{nmp}^{X}}} \\
        &\gamma^U (n, k, \lambda, \varepsilon') = \frac{1}{2 + 2 \frac{A^2 G}{(n+k)^2}} \left\{ \frac{(1 - 2\lambda) A G}{n + k} + \sqrt{\frac{A^2 G^2}{(n+k)^2} + 4\lambda(1 - \lambda) G} \right\}\\
        &A = \max \{n, k\}, \hspace{2cm} G = \frac{n + k}{nk} \log_e \frac{n + k}{2 \pi nk \lambda (1 - \lambda) \varepsilon'^2}
    \end{aligned}
\end{equation}
$\lambda_{EC}$ is the estimation on the known leakage of information from the error correction process,
\begin{equation}
\lambda_{EC}=\left[ n_R^Z (1 -E_Z) - F^{-1} \left( \varepsilon_{cor} \cdot \left( 1 + \frac{1}{\sqrt{n_R^Z}} \right) ; n_R^Z, 1 -E_Z \right) - 1 \right]
\end{equation}
where $E_{Z}=\frac{M_{R}^{Z}}{N_{R}^{Z}}$ is bit error rate of received Z basis count and $F^{-1} \left( \varepsilon_{cor} \left( 1 + \frac{1}{\sqrt{n_R^Z}} \right) ; n_R^Z, 1 -E_Z \right) $ is the inverse of the cumulative distribution function of the binomial distribution. The simulation parameter is displayed in the following Table.\ref{parameter}.

\begin{table}[ht]
\caption{System Parameters}
  \label{tab:sim_parameters}
  \centering
\begin{tabular}{ccc}
\hline
Description & Parameter & Value \\
\hline
Repetition rate & $f_{rep}$ & 75.947 MHz \\
Average photon number per pulse before the quantum channel & $\langle n \rangle$ & $2.89\times10^{-3}$ \\
Second-order correlation & $g^{(2)}$ & $0.85\%$ \\
Transmission efficiency of encoder and decoder & $\eta_{A}$, $\eta_{B}$  & $10.11\%$, $41.7\%$ \\
Z-basis choice (Encoder) & $p_{Z}^{A}$ & $11/16$ \\ 
X-basis choice (Encoder) & $p_{X}^{A}$ & $5/16$ \\
Z- and X- basis choice (Decoder) & $p_{X}^{B}$ & $1/2$ \\
Misalignment probability of Z-basis & $p_{misZ}$ & 1\% \\
Misalignment probability of X-basis & $p_{misX}$ & 2\% \\
Averaged fibre-spool loss & $\alpha$ & \SI{0.1956}{\dB\per\kilo\meter} \\
Detector efficiency & $ \eta_D $ & $74\%$ \\
Dead time & $\tau_{dt}$ & 35.8 ns \\
Time window of one bit & $\tau_{W}$ & 4.3 ns \\
Dark count probability & $p_{dc}$ & $1.33 \times 10^{-6}$ \\
Parameter estimation failure probability & $\epsilon_{PE}$ & $\nicefrac{2 \times 10^{-10}}{3}$ \\
Error correction failure probability & $\epsilon_{EC}$ & $\nicefrac{10^{-10}}{6}$ \\
Privacy amplification failure probability & $\epsilon_{PA}$  & $\nicefrac{10^{-10}}{6}$ \\
Error verification failure probability & $\epsilon_{cor}$ & $10^{-15}$ \\
\hline
\end{tabular}
\label{parameter}
\end{table}

\section*{Acknowledgements}
The authors thank Alessandro Fedrizzi and Frederik Brooke Barnes for the fruitful discussion about SKR simulation, Johann Dzeik for helping with the 3D-printing of encoder and decoder container, and Jialiang Wang for the experimental assistance. The authors gratefully acknowledge the funding by the German Federal Ministry of Education and Research (BMBF) within the project QR.X ({16KISQ013} and 16KISQ015), QR.N (16KIS2188 and 16KIS2207), SQuaD (16KISQ117) and SemIQON (13N16291), and the European Research Council (MiNet GA101043851). We thank the project EQSOTIC within the QuantERA II Programme that has received funding support from the European Union’s Horizon 2020 research and innovation programme under the Grant Agreement No. 101017733, and BMBF (No. 16KIS2060K), the Deutsche Forschungsgemeinschaft (DFG, German Research Foundation) via the Project 469373712, GRK2642, InterSync (GZ: INST 187/880-1 AOBJ: 683478), and under Germany’s Excellence Strategy (EXC-2123) Quantum Frontiers (390837967).

\section*{Data availability}
The data that support the plots within this paper and other findings of this study are available from the corresponding author upon reasonable request.

\section*{Conflict of interest}
Fei Ding serves as an Editor for the Journal. No other author has reported any competing interests.

\section*{Contributions}
J.P. Wang built encoder and decoder setup and carried out the QKD experiment (CRediT: Investigation),  with help of J. Hanel and X.Y Cao (CRediT: Methodology). Z.H. Jiang and J.Z. Yang implemented the optical characterization of the quantum dot sample (CRediT: Validation), with preliminary support from M. Jetter and R. Joos (CRediT: Resources). E.P. Rugeramigabo provided support
with instrumentation and optical experiments (CRediT: Resources). J.P. Wang performed the data analysis (CRediT: Software). J.P. Wang, J.Yang and F.Ding wrote the manuscript (CRediT: Formal analysis, Writing),  with the help of S.L. Portalupi, M. Zopf and the other co-authors (CRediT: Project administration, Writing). F. Ding, M. Zopf and J.Z. Yang conceived and supervised the project (CRediT: Funding acquisition, Conceptualization, Supervision), with support from S.L. Portalupi and P. Michler (CRediT: Validation, Writing).

\bibliography{references.bib}
\end{document}